%% file: root1.tex
\documentclass{article}
\usepackage{setspace}
\title {\bf Duality and cosmological compactification of superstrings with unbroken supersymmetry}
\author{H.K.Jassal\thanks{E--mail : hkj@ducos.ernet.in},
A.Mukherjee\thanks{E--mail : am@ducos.ernet.in} and
R.P.Saxena\thanks{E--mail : rps@ducos.ernet.in} \\
	{\em Department of Physics and Astrophysics,} \\
	{\em University of Delhi, Delhi-110 007, India} 
	}

\begin {document}

\maketitle
\begin {center}
\Large {\bf Abstract}
\end{center}

\large

   \input {abstract.tex}

\vfill
\eject
\large

\baselineskip=15pt
\begin{section} {Introduction}
   \input {intro1.tex}

   \end {section}

\begin{section} {Compactification}
   \input {compact.tex}

   \end {section}

\begin{section} {Dual Equations}
   \input {dual1.tex}

   \end {section}

\begin{section} {Summary and Conclusions}
   \input {conclude.tex}

   \end {section}

\begin{section}*{Acknowledgements}
   \input {ack.tex}
   \end {section}
\vfill
\eject
\bibliography{plain}

\vfill
\eject
\begin{section}*{Figure Captions}
      \input{figure.tex}
      \end{section}
\input{asmr.tex}
\begin{center}
FIG. 1
\end{center}
\vfill
\eject
\input{asmdil.tex}
\begin{center}
FIG. 2
\end{center}
\vfill
\eject
\input{newrl.tex}
\begin{center}
FIG. 3
\end{center}
\vfill
\eject
\input{newdl.tex}
\begin{center}
FIG. 4
\end{center}
\vfill
\eject
\input{rl.tex}
\begin{center}
FIG. 5
\end{center}
\vfill
\eject
\input{chil.tex}
\begin{center}
FIG. 6
\end{center}
\vfill
\eject
\input{r1l.tex}
\begin{center}
FIG. 7
\end{center}
\vfill
\eject
\input{chi1l.tex}
\begin{center}
FIG. 8
\end{center}
\vfill
\eject
\input{dill.tex}
\begin{center}
FIG. 9
\end{center}
\vfill
\eject
\input{dil1l.tex}
\begin{center}
FIG. 10
\end{center}
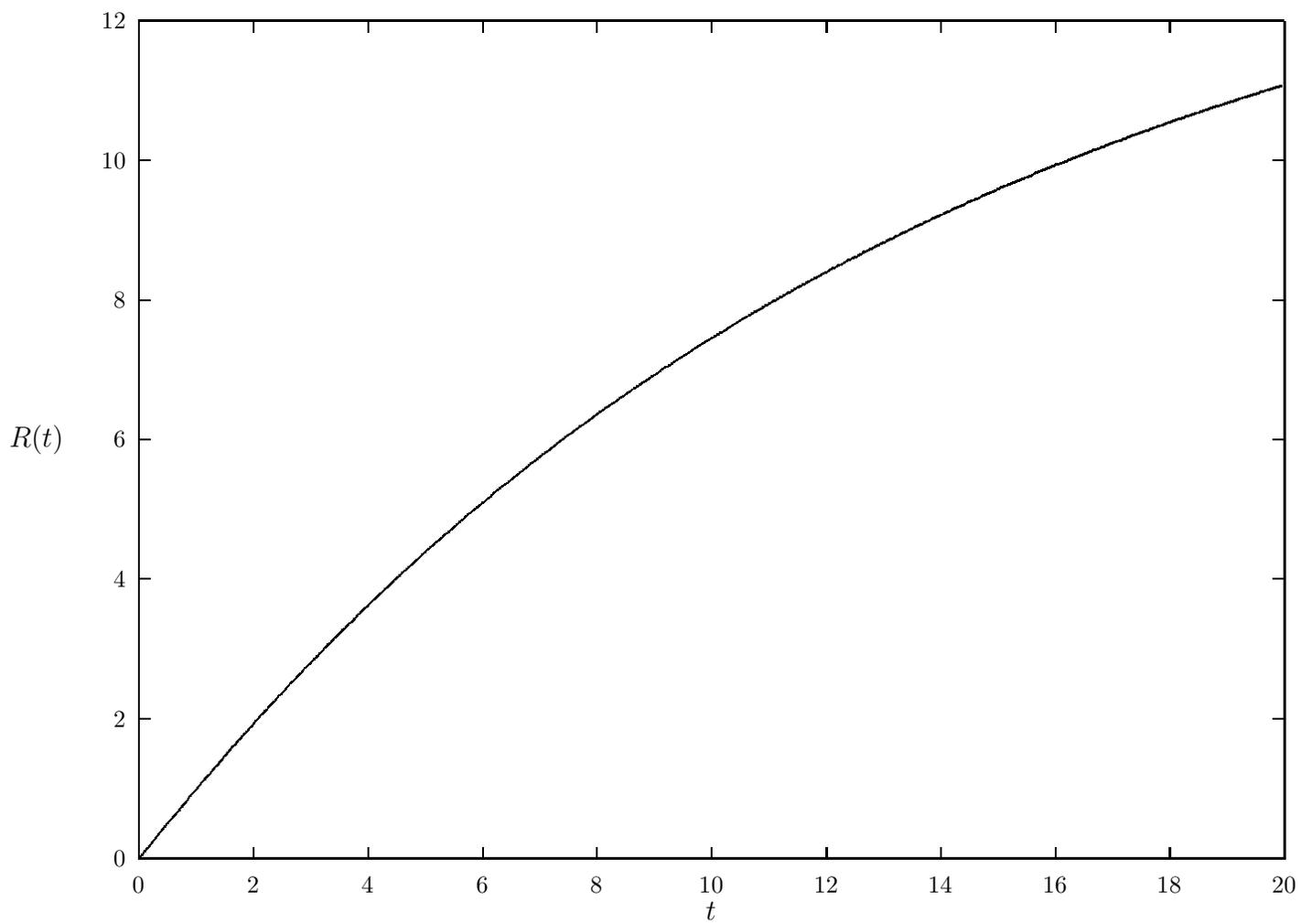
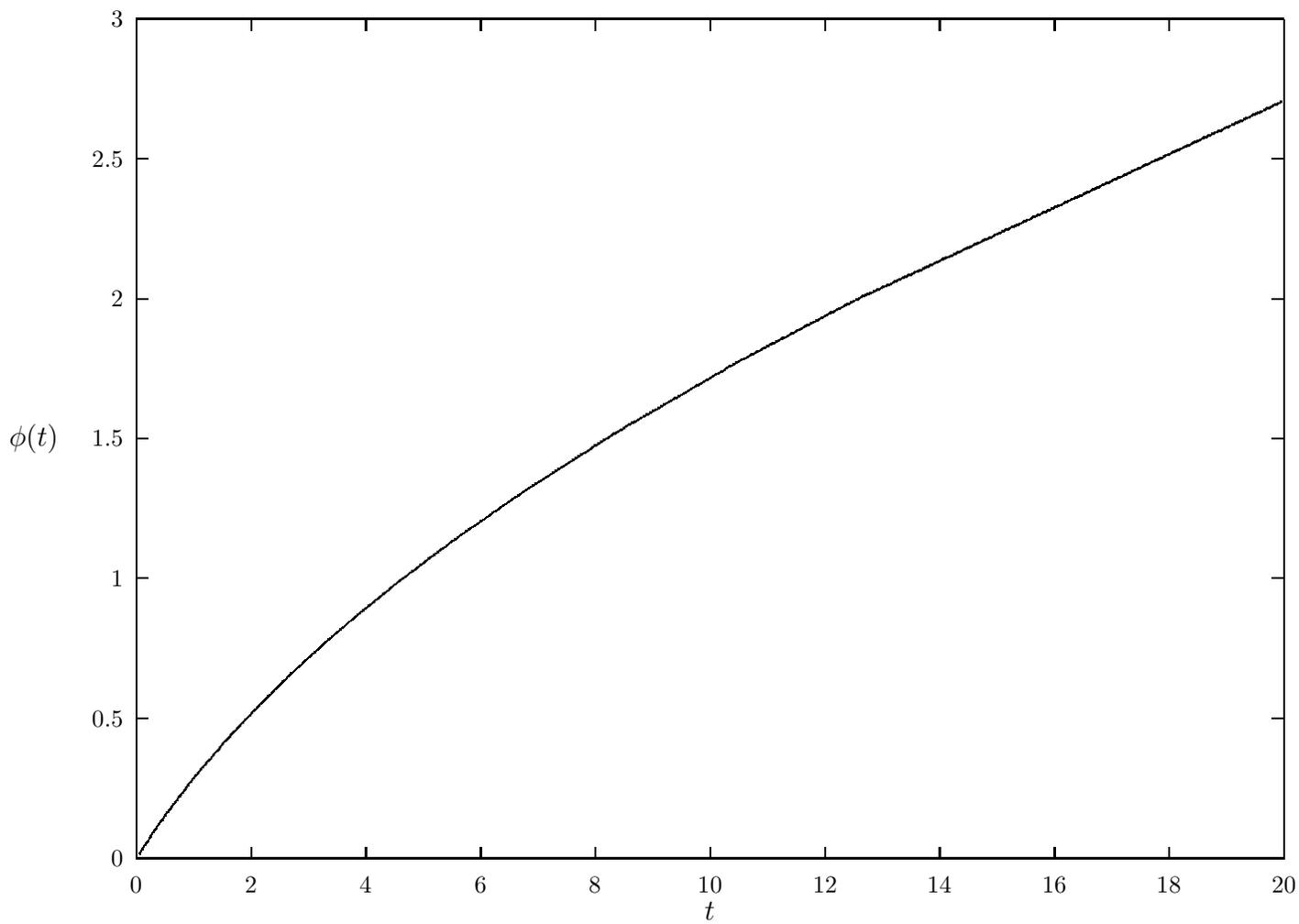
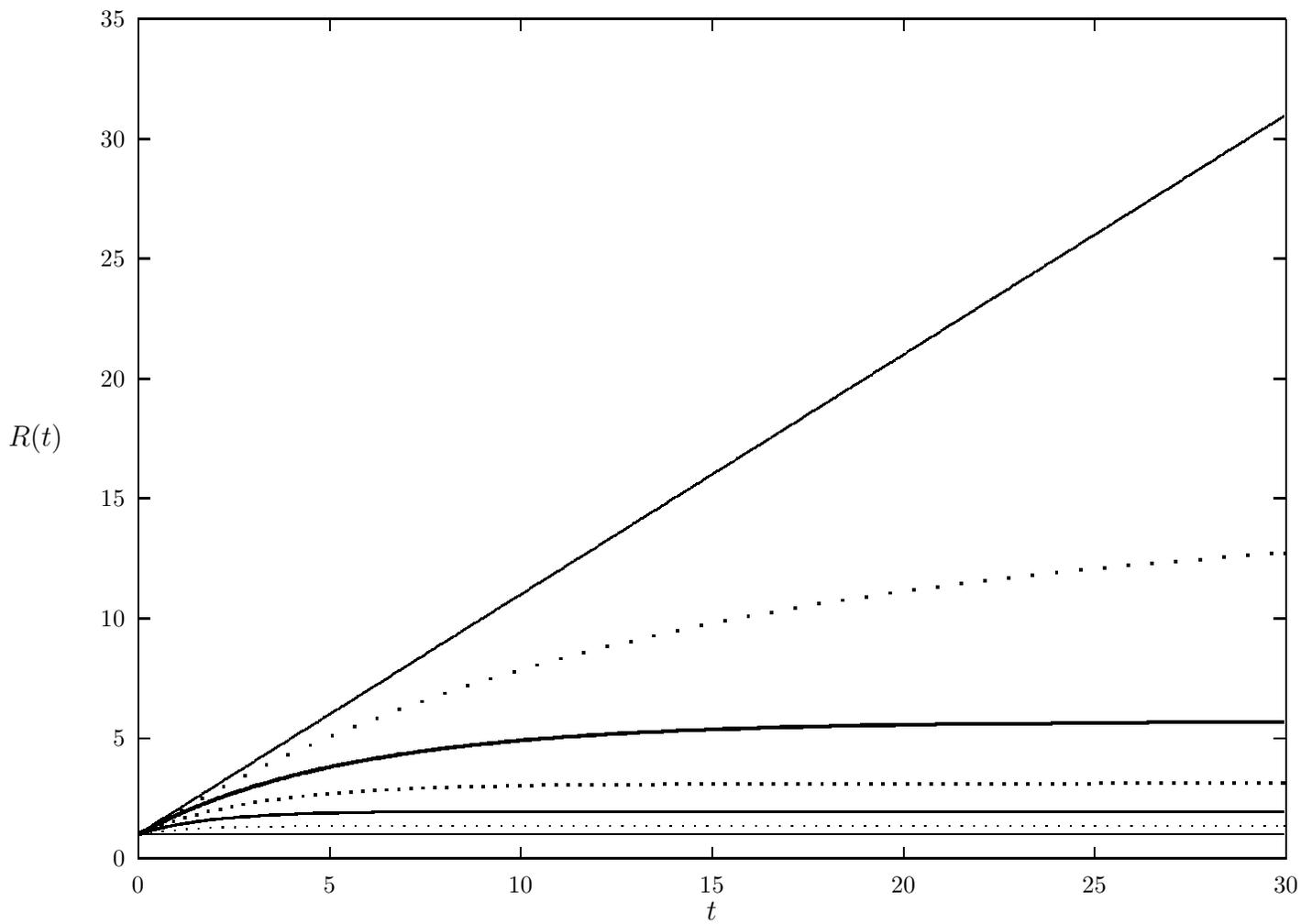
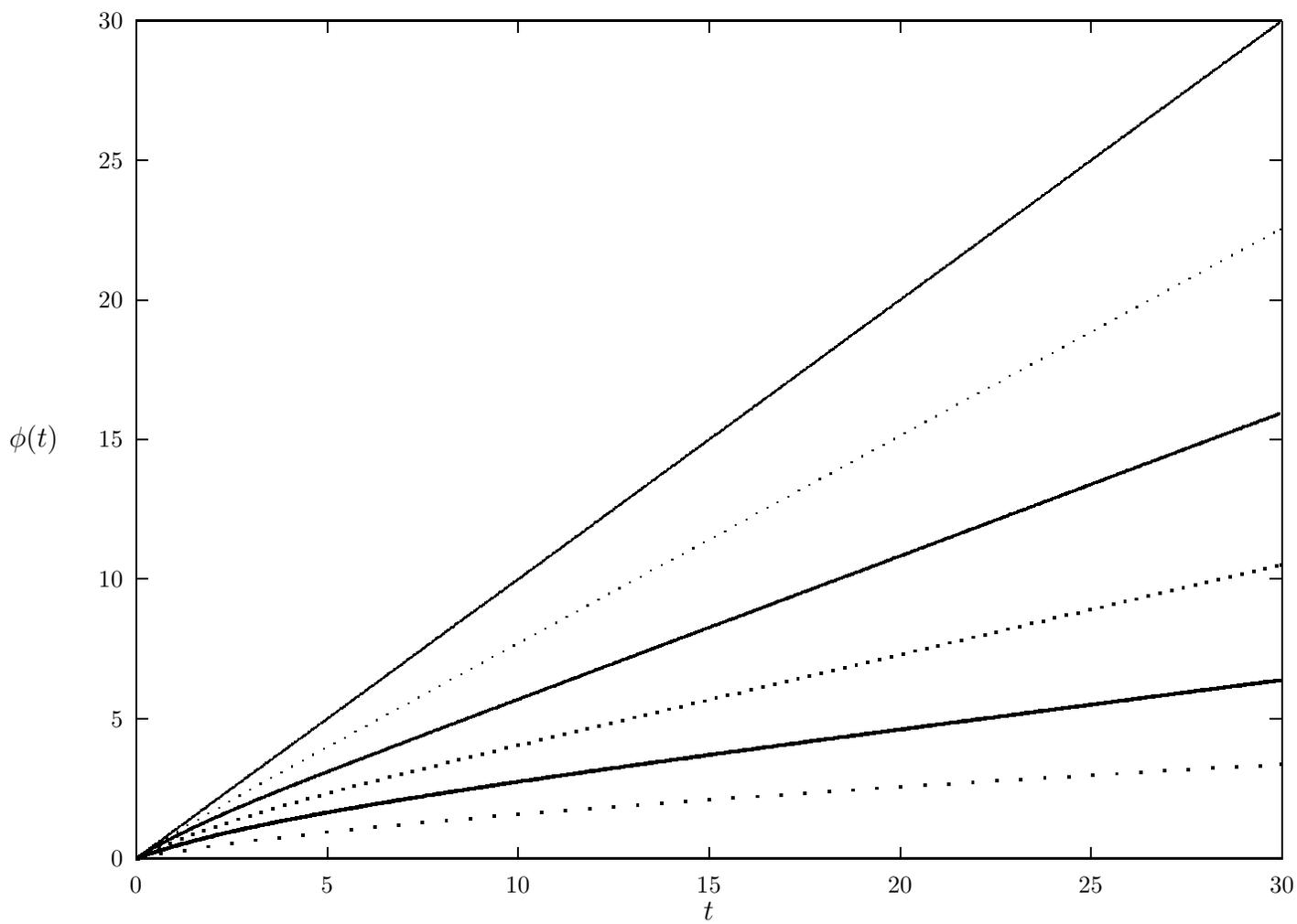
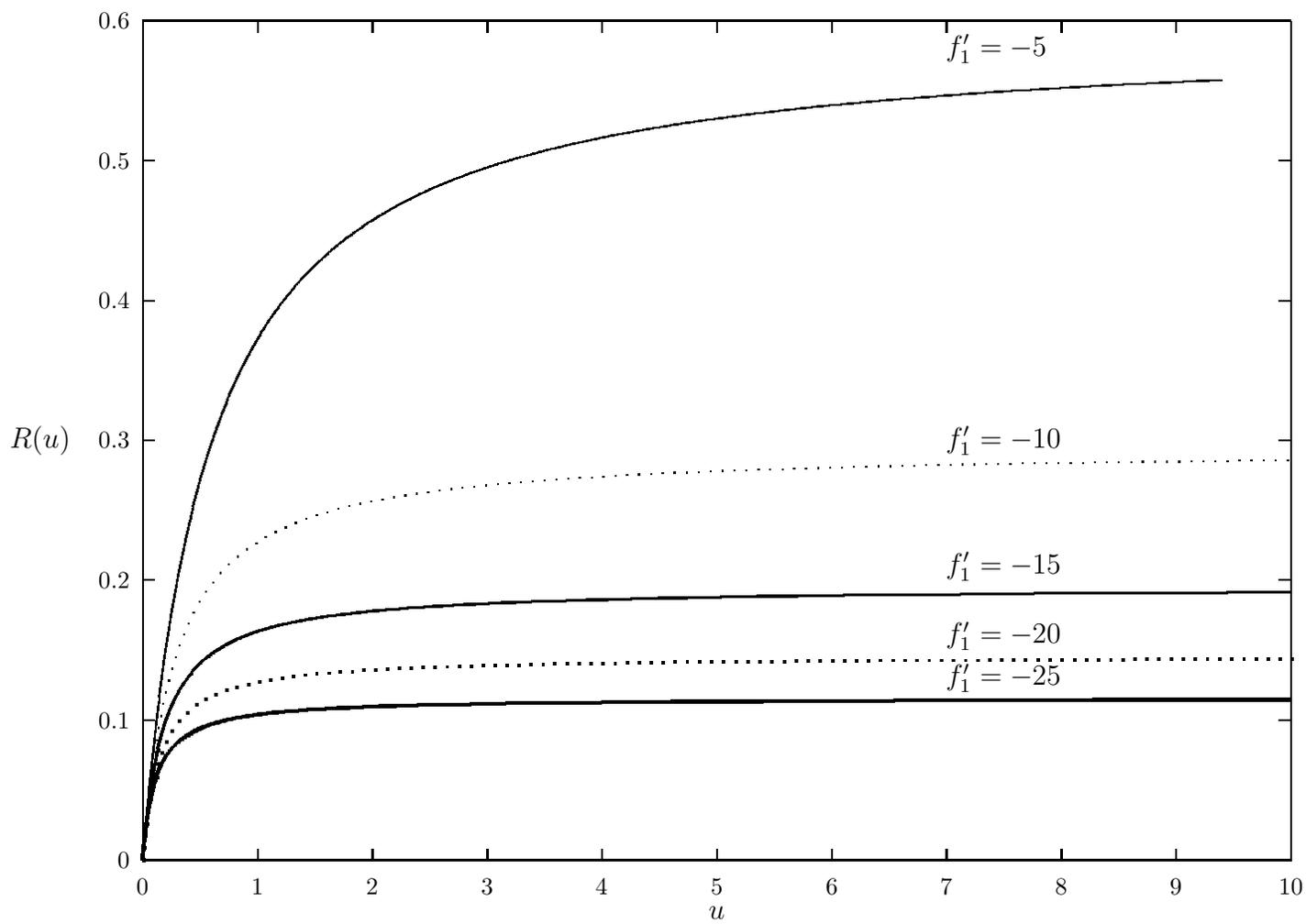
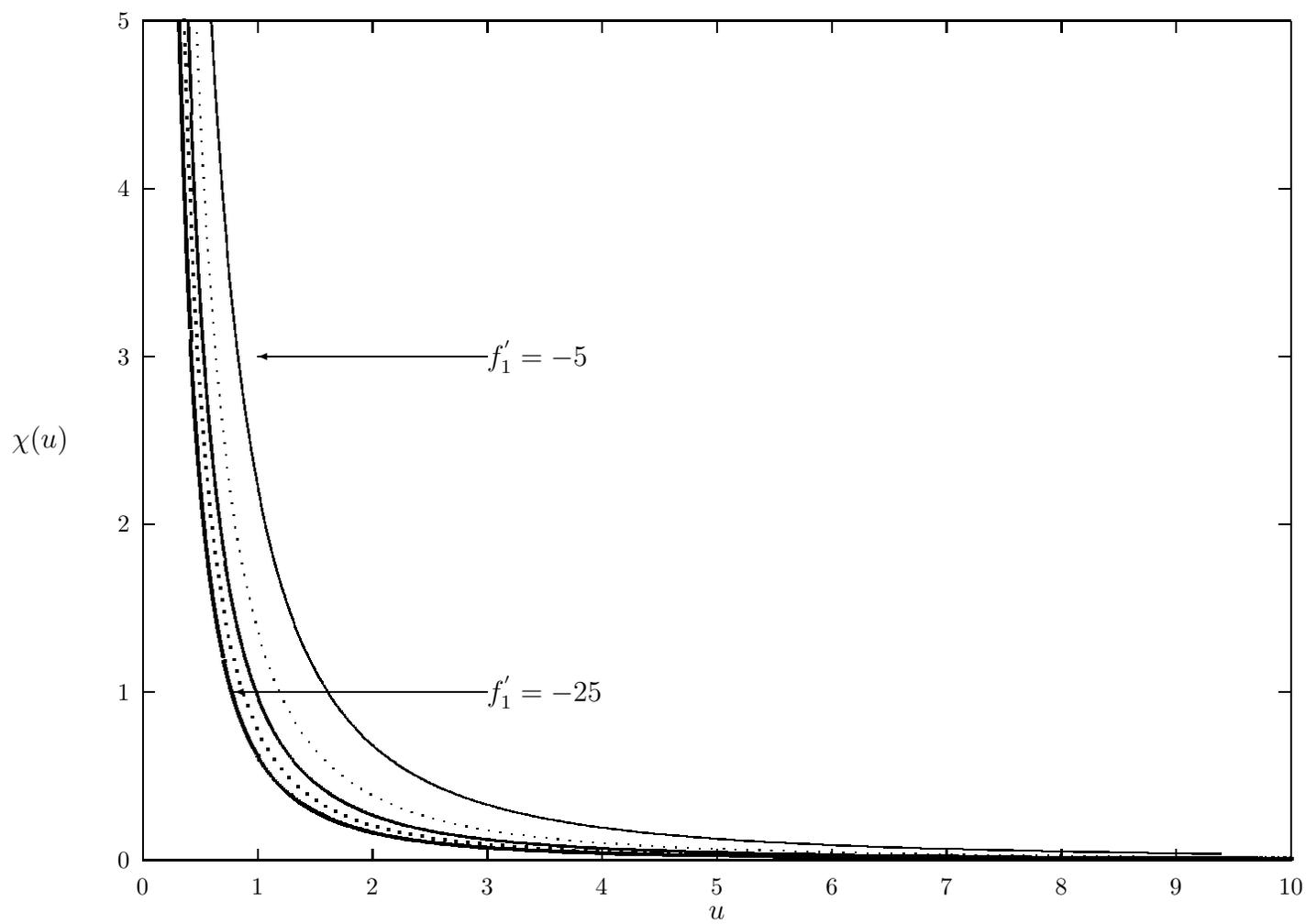
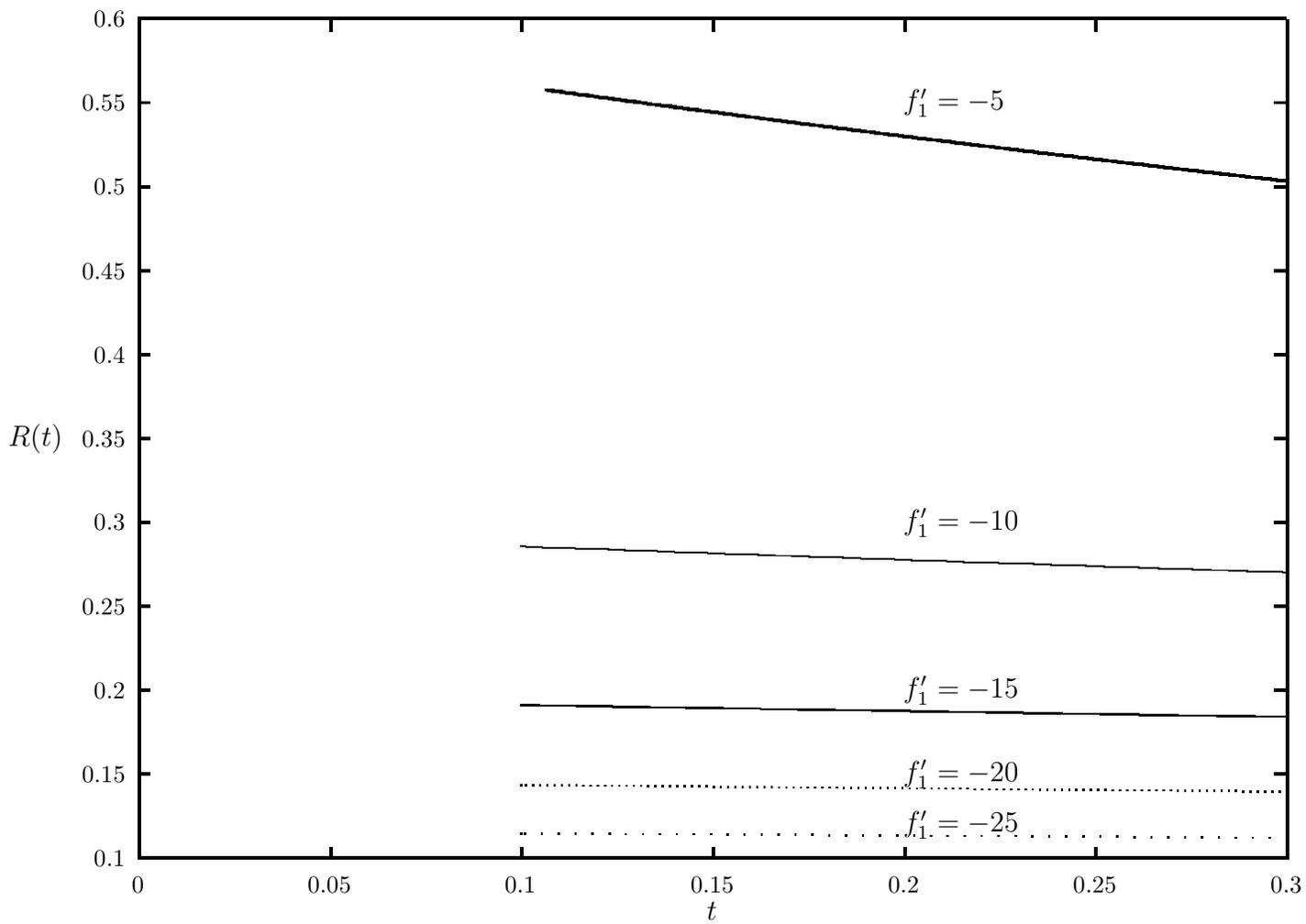
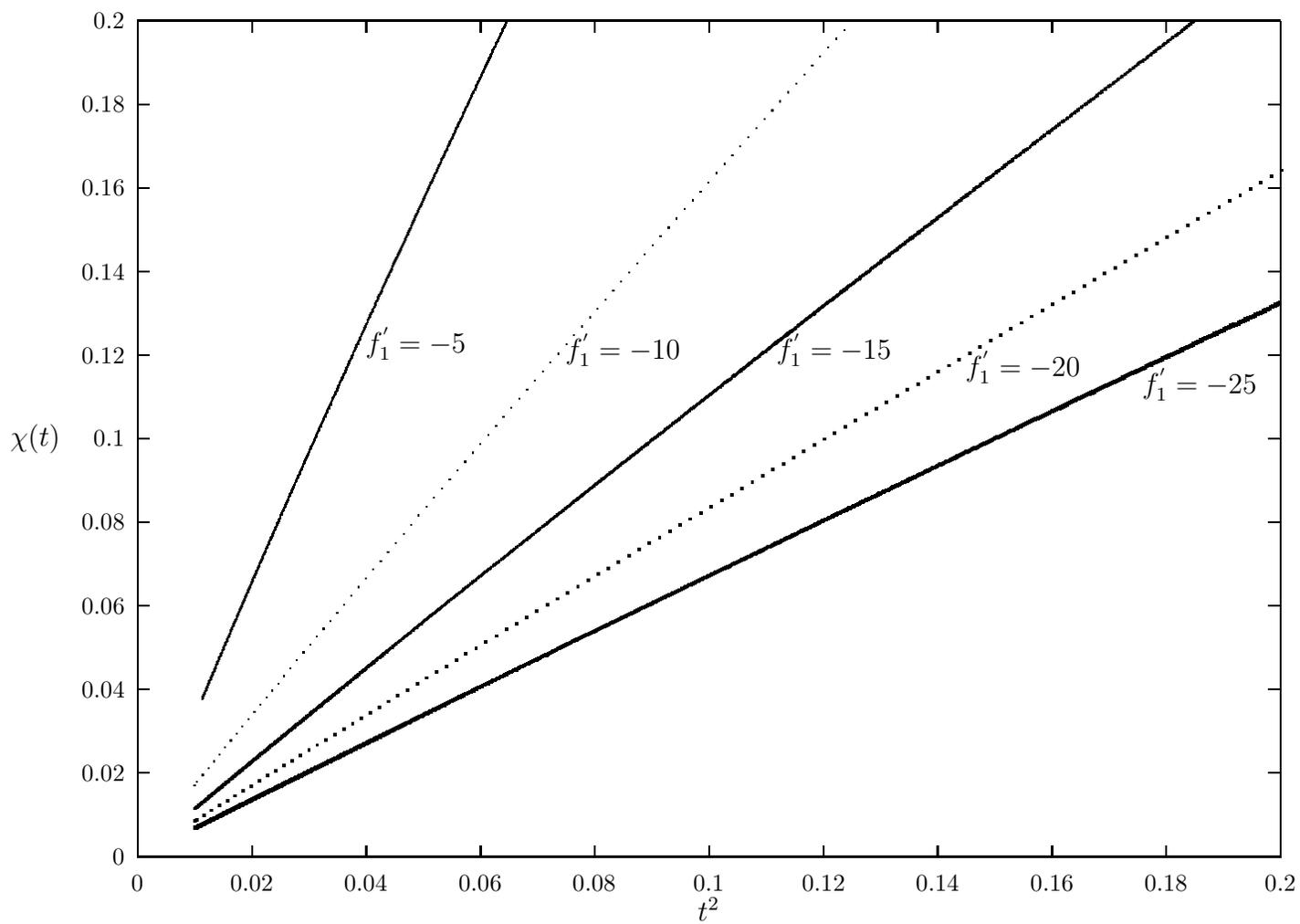
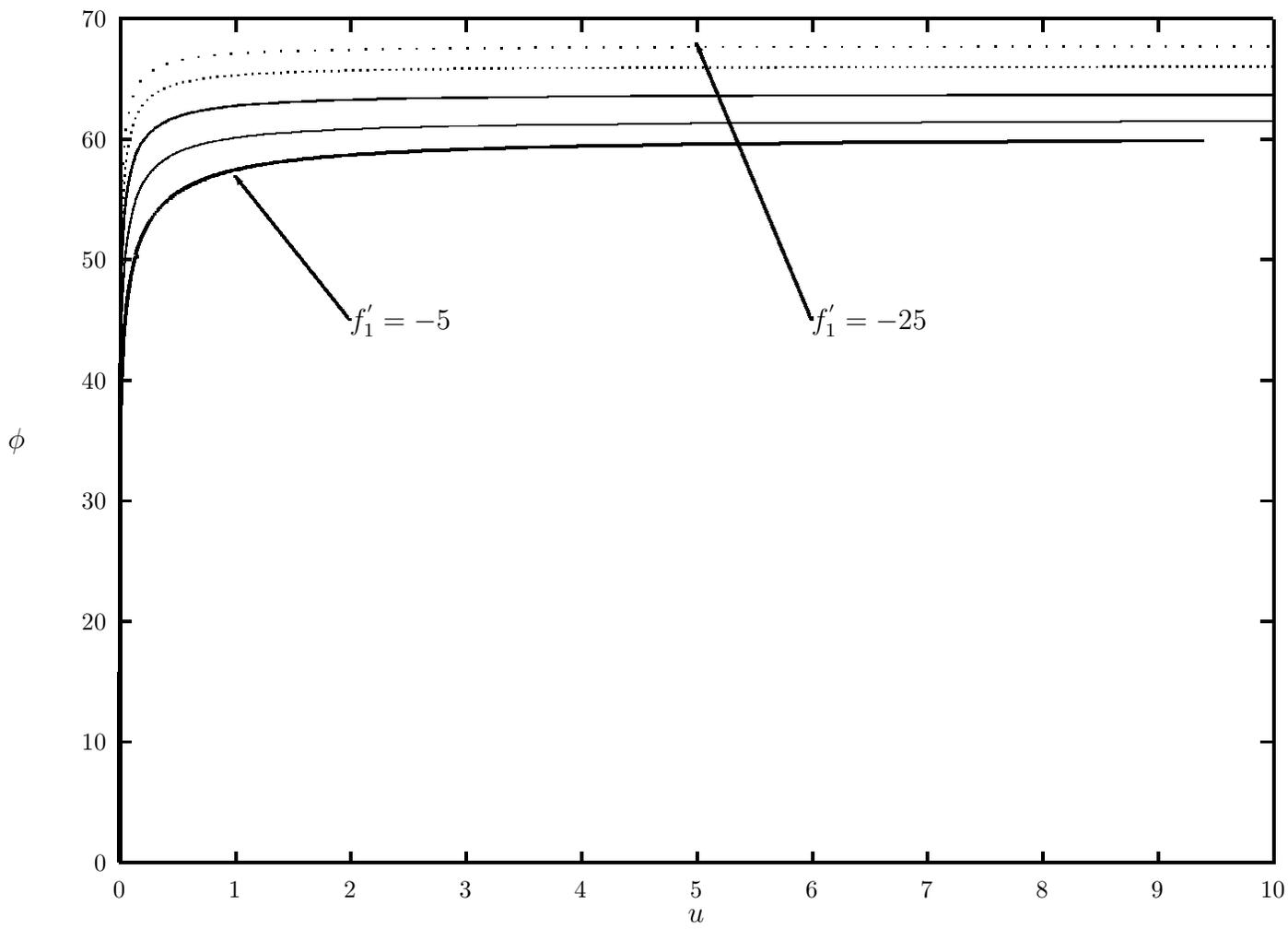
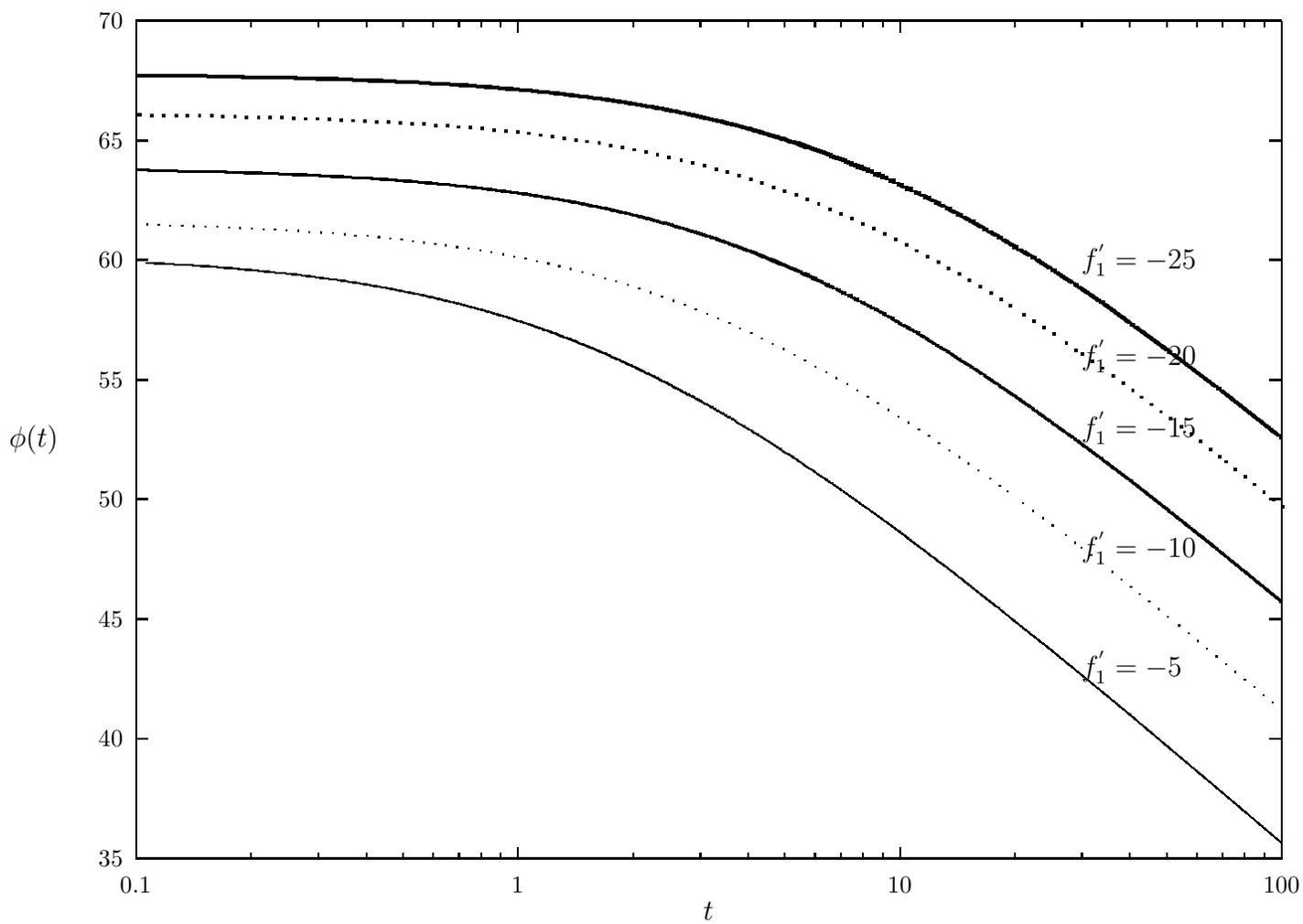
\end{document}

%% file: abstract.tex
%
%
\noindent The cosmological compactification of $D=10$ , $N=1$
supergravity--super--Yang--Mills theory obtained from superstring
theory is studied.
The constraint of unbroken $N=1$ supersymmetry is
imposed.
A duality transformation is performed on the resulting
consistency conditions.
The original equations as well as the transformed equations 
are  solved numerically to obtain new
configurations with a nontrivial scale factor and a dynamical
dilaton.
It is shown that various classes of solutions are possible,  which include 
cosmological solutions with no initial singularity.

%% file: intro1.tex
%
%

\indent String theory \cite{Green}, at present, is the most promising
candidate for a unified theory of gravity with other forces of nature.
Therefore, there is considerable interest in investigating string
effects on  the evolution of the universe.
In principle, string theory is expected to explain the curved nature
of spacetime, provide a consistent description
of the very early history of the universe and generate corrections to
the cosmological evolution equations which are in agreement with 
current observations.

\vspace{0.5cm}
String theory is defined on a D--dimensional manifold (D$=$26 
for bosonic strings and D$=$10 for supersymmetric strings).
Thus to make a connection  with physical four--dimensional -- in
particular, cosmologically relevant -- geometries, the extra
D$-$4 dimensions have to be compactified. 
Specifically, we require the D--dimensional manifold to be a direct
product of the four--dimensional manifold and the (D$-$4)--dimensional
compact internal space whose physical dimensions are of the order of the
Planck scale.

\vspace{0.5cm}
Candelas {\em et al}. \cite{Candelas} considered the compactification
of ten--dimen- \\ sional supergravity coupled to super--Yang--Mills theory,
assuming unbroken  supersymmetry and maximal symmetry in
four dimensions.      
It was found that consistent compactification was possible only for
Minkowskian geometry. 
The condition of maximal symmetry in four dimensions is
restrictive, and relaxing this condition allows compactification to
curved spacetime \cite{Strominger}. 
Explicit solutions to the consistency conditions for unbroken
supersymmetry were obtained \cite{Kumar} within the ansatz of
spherical symmetry for three--dimen- \\ sional space.
It was found that the scale factor has a Milne type evolution, with a
constant dilaton field in four dimensions.  
In general, if a dynamical dilaton $\phi$ and a non--vanishing
antisymmetric field strength tensor $H_{\mu\nu\rho}$  are considered,
nontrivial evolution \cite{MMS} of the scale factor and the dilaton
field is obtained in  an open Friedmann--Robertson--Walker (FRW)
universe. 

\vspace{0.5cm}
Recent work in string cosmology has focussed on solutions of the equations
of motion obtained from the low--energy effective  action \cite{Tree},
i.e., the action  describing  the dynamics of massless modes (the
metric $g_{\mu\nu}$, the 
dilaton $\phi$ and the antisymmetric field $B_{\mu\nu}$) of the
string. 
It was pointed out by Veneziano \cite{Veneziano} that some of the duality
symmetries of string theory are manifested as symmetries of the
low-energy field equations.  
In particular, he emphasised the importance of scale factor duality
(SFD), which relates physically inequivalent solutions of the string
modified Einstein--Friedmann equations. 
For an expanding universe with increasing scale factor $R(t)$, we have a
dual scenario with $R(t)\rightarrow R(t)^{-1}$ which describes a
contracting universe.
Although the ``proof'' of SFD \cite{Veneziano} is valid only for spatially flat
geometries, it seems reasonable to assume that it is more generally true (see,
e.g., \cite{Ossa}).
Several attempts have been made \cite{Brustein, OliveNPB, OlivePLB,
Jmu, Sukanta} to use SFD to solve the graceful exit problem of  
inflationary cosmology. 
This typically requires introducing by hand an {\it ad-hoc} dilaton
potential. 
At a deeper level, scale factor duality offers a way to solve the
initial singularity problem of big-bang cosmologies \cite{Jmu, PREBB}.

\vspace{0.5cm}
The above considerations, however, may be insufficient for a truly
``stringy'' description of cosmological evolution.
For instance, more theoretical support is required to justify the form
of  the dilaton potential.
For our purpose, we do not assume any explicit form for the effective
action. 
Taking an alternative approach, we work within the framework of
supersymmetric compactification.
We are encouraged in this by the fact that unbroken supersymmetry in
four dimensions, at some energy scale high compared to the
electroweak scale, seems to be required in most viable theoretical
scenarios  \cite{Ross} for the unification of forces. Assuming that 
scale factor duality
holds for spatially non-flat geometries,  
we explore the consequences   
on the consistency conditions for compactification with
unbroken supersymmetry. 
In particular, we find out the four--dimensional geometries consistent
with the duality transformed equations.
It is shown that different classes of cosmological solutions are
allowed, including those with no initial singularity.
Since  we do not assume any arbitrary potential, these considerations
may, in general, be applicable to a wider range of cosmological
evolution possibilities.

\vspace{0.5cm}
The paper is organised as follows.
In section II  we consider the compactification of ten--dimensional
supergravity--super--Yang--Mills theory with $N=1$ supersymmetry.
It is shown that compactification with unbroken supersymmetry admits
different classes of nontrivial cosmological solutions. 
Section III is devoted to the solutions of equations which are
transformed under the duality transformations. 
We show that an entirely different class of cosmological solutions is
obtained. 
Section IV contains some concluding remarks.

%% file: compact.tex
%
%
\newcommand{\del}{\mbox{$\bigtriangledown$}} 
\indent We consider the compactification of ten--dimensional supergravity
coupled to super--Yang--Mills theory \cite{Chapline}, i.e., the
infinite tension limit of superstring theory.  
The 10--dimensional manifold is of the form
$M_{D}=\mathcal{M}_{4}\otimes\mathcal{K}_{6}$ where $\mathcal{M}_{4}$
is the four--dimensional spacetime and $\mathcal{K}_{6}$ 
is the six--dimensional compact internal manifold whose  physical 
dimensions are of the order of the Planck scale.
For simplicity, we assume the FRW form for the metric in  
four--dimensional spacetime.

\vspace{0.5cm}
The fermionic fields in the supermultiplet are assumed to have
vanishing background values.
The supersymmetric transformations for the fermi fields for some
Majorana--Weyl spinor $\epsilon$  are 

\begin{equation}
\delta\varphi_\mu = \del_\mu\epsilon + \frac{1}{48} e^{2
\phi} (\gamma_{\mu}  H_{\rho\nu\sigma} \gamma^{\rho\nu\sigma} - 
12 H_{\mu\rho\nu} \gamma^{\rho\nu} + \gamma_{\mu} \gamma_{5}\otimes H)
\epsilon
\end{equation}
\vskip 0.2 cm
\begin{equation}
\delta\lambda = \gamma^{\mu}(\del_\mu\phi)\epsilon +
 \frac{1}{24}e^{2\phi} H_{\rho\nu\sigma} \gamma^{\rho\nu\sigma}\epsilon
 + \gamma_{5} \otimes [ \gamma^{m}(\del_{m}\phi) +
\frac{1}{24}e^{2\phi} H] \epsilon
\end{equation}

\vspace{0.5cm}
\noindent Here $\gamma^{\rho\nu}=\frac{1}{2}\gamma^{[\rho}\gamma^{\nu]}$ and
$H \equiv H_{mnp}\gamma^{mnp}$, $H_{\mu\nu\rho}$ being the
field strength tensor for the antisymmetric (torsion) tensor.
The Greek indices represent the four--dimensional spacetime, m,
n... represent the six--dimensional compactified  manifold and M,
N... denote the full ten--dimensional spacetime.

\vspace{0.5cm}
Since we demand spacetime supersymmetry for the ground state,
$\delta\varphi_{\mu}$ and $\delta\lambda$ must vanish for any
arbitrary $\epsilon$. 
The requirement of maximal symmetry in {\it three} dimensions leads us to
the following ansatz for $H_{\mu\nu\rho}$ and the dilaton field $\phi$ 

\begin{eqnarray}
\phi & = & \phi(r,t) \\
H_{\mu\nu\rho} & = & e^{l(r)} \varepsilon_{\mu\nu\rho\sigma} b^{\sigma}
\end{eqnarray}

\vskip 0.2 cm
\noindent where l(r) is any arbitrary function of r and $b^{\sigma}$ is a
constant timelike vector.

\vspace{0.5cm}
	Taking the dilaton field to be a constant in internal space
\cite{Gates}, i.e., $\del_{m}\phi =0$, the requirement  
of unbroken supersymmetry, using equations (1) and (2),  yields 

\begin{eqnarray}
[\del_{\mu},\del_{\lambda}]\epsilon & = &
 \frac{1}{2} [\gamma_{\lambda}(\del_{\mu}\del_{\sigma}\phi)-
\gamma_{\mu}(\del_{\lambda}\del_{\sigma}\phi)]\epsilon \nonumber \\
                 &   &+\frac{1}{4}[\gamma_{\lambda}\gamma^{\sigma}
\gamma_{\mu}\gamma^{\tau}-\gamma_{\mu}\gamma^{\sigma}\gamma_{\lambda}
\gamma^{\tau}](\del_{\sigma}\phi)(\del_{\tau}
\phi)\epsilon \nonumber \\
                 &   & \mbox{}+\frac{1}{4}[\del_{\mu}(2\phi+l)
\varepsilon_{\lambda\rho\nu\sigma}
-\del_{\lambda}(2\phi+l)\varepsilon_{\mu\rho\nu\sigma}]\gamma^{\rho\nu}
b^{\sigma}\epsilon \nonumber \\
                 &   & \mbox{}+\frac{1}{16} e^{4\phi+2l}
[\varepsilon_{\lambda\rho\nu\sigma}\varepsilon_{\mu\delta\eta\tau}-
\varepsilon_{\mu\rho\nu\sigma}
\varepsilon_{\lambda\delta\eta\tau}]
\gamma^{\rho\nu}\gamma^{\delta\eta}b^{\sigma}
b^{\tau}\epsilon  \nonumber
\end{eqnarray}

\vspace{0.5cm}
As in Candelas {\em et al}. \cite{Candelas}, we have
$[\del_{\mu}, \del_{\lambda}]\epsilon = -\frac{1}{4}R_{\mu\lambda\nu\rho}
\gamma^{\nu\rho}\epsilon$, where $R_{\mu\lambda\nu\rho}$ is the Riemann
curvature tensor.
To avoid torsion in four--dimensions, we do not consider the case when
both $\phi$ and $H_{\mu\nu\rho}$ have nontrivial behaviour, i.e.,
either we take the dilaton field to be a constant in four dimensions
or we take $H_{\mu\nu\rho}=0$. 
The above integrability condition gives the Riemann tensor in terms of
fields $\phi(t)$ and $H_{\mu\nu\rho}$, which in turn gives rise to a
set of coupled equations, which are similar to the equations of motion
obtained from the action in scalar--tensor theories of gravity. 
These equations can be solved to find four--dimensional geometries
consistent with the requirement of unbroken supersymmetry.

\vspace{0.5cm}
It was shown  \cite{MMS} that, in the case where the dilaton field is a
constant, i.e., for $\del_{\mu}\phi=0$,  we have

\begin{equation}
\frac{\ddot R}{R}=0
\end{equation}
\vskip 0.2 cm
\begin{equation}
\dot R^{2}+cR^{2}+k=0
\end{equation}.

\vspace{0.5cm}
\noindent Here we have imposed $\del_{\mu}l=0$, and the constant c is
defined by $c=e^{4\phi+2l}$. 
Clearly, the only solution in this case is the Minkowskian geometry. 

\vspace{0.5cm}
If $H_{\mu\nu\rho}=0$, then for a time dependent dilaton field, the
equations take the form  
\cite{MMS} 

\begin{equation}
\frac{\ddot R}{R}-\ddot \phi=0
\end{equation}
\vskip 0.2 cm
\begin{equation}
\dot R^{2} + R^{2}\dot\phi^{2} + k=0
\end{equation}

\vspace{0.5cm}
If $\dot\phi=0$, we again get the Minkowskian solution.
To get consistent nontrivial solutions, $k$, the FRW spatial curvature
factor cannot have any value greater than or equal to zero.
Thus, we have $k=-1$, corresponding to an open universe. 

\vspace{0.5cm}
Numerical solutions of these equations \cite{MMS} show that the dilaton 
field approaches a linear growth in time while the scale factor
grows subluminally (see Fig.(1) and Fig.(2)).
This shows that compactification of 10--dimensional
supergravity with unbroken supersymmetry is consistent with nontrivial
evolution for the scale factor and the dilaton field in
four dimensions.

\vspace{0.5cm}
In addition to the solutions presented in \cite{MMS}, there also
exists a class of one--parameter solutions with the initial conditions
$R(0)=1$, $\dot R(0)=\alpha$, $\dot \phi(0)=\beta$, with
$\alpha^{2}+\beta^{2}=1$. 
The numerical solutions with these initial conditions are shown in
Fig.(3) and Fig.(4).
It is clear that the consistency conditions allow non--singular
solutions.
Note that again the dilaton field approaches a linear growth but the
scale factor approaches a constant for large time. 

\vspace{0.5cm}
In general, the equations can be reparametrised as

\begin{eqnarray}
\dot R(t)  & = &  \cos\theta(t) \nonumber \\
&& \nonumber \\
R(t) \dot\phi(t)  & = &  \sin\theta(t) \nonumber
\end{eqnarray}

\vspace{0.5cm}
which give the following solutions

\begin{eqnarray}
R(t) & = & A  e^{\theta(t)} \sin\theta(t) \nonumber \\
&& \nonumber \\ 
\dot\phi(t) & = & \frac{1}{A}  e^{-\theta(t)} \nonumber 
\end{eqnarray}

\vspace{0.5cm}
Interestingly, the non--singular solutions obtained above are a
special case of this more general parameterization, i.e., that
corresponding to

\begin{eqnarray}
Ae^{\theta(0)}\sin\theta(0) &=& 1 \nonumber
\end{eqnarray}

%% file: dual1.tex
\indent We now transform the consistency conditions (equations (7) and (8))
under SFD, i.e., the scale factor is replaced by its inverse, namely 

\begin{equation}
R(t)\rightarrow\frac{1}{R(t)} \nonumber
\end{equation} 

\vspace{0.5cm}
It was shown \cite{Veneziano} that the string--modified
Einstein--Friedmann equations are invariant under the above
transformation if the time--dependent dilaton field $\phi(t)$
transforms nontrivially  as 
 
\begin{equation}
\phi(t)\rightarrow\phi(t)-\ln \mid g_{ii}\mid \nonumber
\end{equation}

\vspace{0.5cm}
\noindent where the spatial indices {\it i} are summed over.
 For an FRW metric with scale factor R(t), the transformation for the 
dilaton field translates to

\begin{equation}
\phi(t)\rightarrow\phi(t)- 6 \ln R(t).
\end{equation} 

\vspace{0.5cm}
We make a working hypothesis that the  transformation given by eqs.(9)
and (11) describes a duality symmetry of the  theory  even though we
cannot establish it in the absence of an action.
We adopt the point of view that, just as the consistency conditions
(eqs. (7) and (8)) lead to cosmological background configurations for
the original theory, the transformed equations describe  cosmological
configurations for the dual theory.
Applying the duality transformations to equations (7) and (8), we get

\begin{equation}
5 \frac{\ddot R}{R} -  4 \left(\frac{\dot R}{R}\right)^{2} - \ddot \phi = 0
\end{equation}
\vskip 0.2 cm
\begin{equation}
37 \dot R^{2} + R^{2}\dot \phi^{2} - 12 R \dot R\dot \phi + k R^{4} = 0
\end{equation}

\vspace{0.5cm}
Changing to a convenient variable $u=\frac{1}{t}$, the equations can
be rewritten as  

\begin{equation}
5 u \frac{d^{2}R}{du^{2}}+10 \frac{dR}{du}-
\frac{4}{R}\left(\frac{dR}{du}\right)^{2} - Ru\frac{d^{2}\phi}{du^{2}}
-2Ru\frac{d\phi}{du}=0
\end{equation}
\vskip 0.2 cm
\begin{equation}
37u^{4}\left(\frac{dR}{du}\right)^{2}+R^{2}u^{4}\left(\frac{d\phi}{du}\right)^{2}-12Ru^{4}
\frac{dR}{du}\frac{d\phi}{du}+kR^{4}=0
\end{equation}

\vspace{0.5cm}
Again, for consistent and nontrivial solutions, we can only have open
FRW configurations, i.e., $k=-1$.

\vspace{0.5cm}
For small u, the scale factor R and $\chi\equiv \frac{d\phi}{du}$
have the following leading order behaviour

\begin{eqnarray}
R  \stackrel{u\rightarrow0}{\longrightarrow}  u \nonumber \\
&& \nonumber \\
\chi \stackrel{u\rightarrow0}{\longrightarrow} \frac{c}{u} \nonumber
\end{eqnarray}

\vspace{0.5cm}
It is convenient to make the transformation of variable given by

\begin{eqnarray}
R &=& uf_{1}(u)  \\
&& \nonumber \\
\chi &=& \frac{c}{u}f_{2}(u) 
\end{eqnarray}

\vspace{0.5cm}
where $f_{1}(0)=1$, $f_{2}(0)=1$ and $f_{1}^{\prime}(0)$ is a
free parameter.
It is interesting to note that the value of the
constant $c$ is fixed by the equations themselves; on substituting
equations (16) and (17) in equations (14) and (15), it turns out that $c=6$.

\vspace{0.5cm}
In terms of the new  variables $f_{1}$ and $f_{2}$, equations (14) and
(15) take the form

\begin{equation}
5 u^{2}f_{1}^{\prime\prime}+10 (uf_{1}^{\prime}+f_{1})-\frac{4}{f_{1}}(uf_{1}^{\prime}+f_{1})^{2}
-6uf_{1}f_{2}^{\prime}-6f_{1}f_{2}=0
\end{equation}
\vskip 0.2 cm
\begin{equation}
37(uf_{1}^{\prime}+f_{1})^{2}+36f_{1}^{2}f_{2}^{2}-72f_{1}f_{2}(uf_{1}^{\prime}+f_{1})-f_{1}^{4}=0
\end{equation}

\vspace{0.5cm}
Numerical solutions of these equations, showing the evolution  of R
and $\chi$ as functions of $u$, are shown in Fig.(5) and Fig.(6).


\vspace{0.5cm}
The leading order analysis of equations (14) and (15) shows the
following asymptotic behaviour 
for R and $\chi$ 

\begin{eqnarray}
R  & \stackrel{u\rightarrow\infty}{\longrightarrow} & A + \frac{B}{u}
\nonumber \\
&& \nonumber \\
\chi & \stackrel{u\rightarrow\infty}{\longrightarrow} & \frac{C}{u^{2}} \nonumber
\end{eqnarray}

\vspace{0.5cm}
\noindent This can be considered as a consistency check for the numerical
solutions of the duality transformed equations. 

\vspace{0.5cm}
If, instead of $u$, we consider the evolution with respect to
$t(=\frac{1}{u})$ the solutions  behave as shown in  Fig.(7) and
Fig.(8). 
Clearly, the scale factor  decreases linearly with $t$ while $\chi$
is proportional to  $t^{2}$, as expected from the argument above.
Notice that for $t\rightarrow 0$ the scale factor approaches a
finite value; thus these solutions have no initial singularity.
From the behaviour of $\chi$ it is clear that the dilaton field
approaches a constant for large $u$ (see Fig.(9)).
We can see this more explicitly by plotting $\phi$ against $t$ (see
Fig.(10)), i.e., the dilaton field is constant for small $t$ while  it
vanishes as $t$ becomes large.  
This shows an entirely different evolution 
behaviour from the one we had obtained in the case of the original
equations.

%% file: conclude.tex
We have considered the compactification of ten dimensional
supergravity coupled to super--Yang--Mills theory and obtained
numerical solutions to the consistency conditions for unbroken
supersymmetry in four dimensions. 
It is clear that supersymmetric compactification allows  nontrivial
geometries which are of cosmological interest.
In the original form, the solutions can describe the evolution of an
FRW universe at late times.
The dual solutions, on the other hand, correspond to interesting
nonsingular cosmologies at early times.
These features are similar to those found by other authors who solve
the low- energy equations of motion.
These results are a reflection of the fact that scale factor
duality does not 
simply reparametrise the equations, but relates two different physical
domains of the theory.

\vspace{0.5cm}
In the above, ``early times'' refers to the time scale at which
supersymmetry is unbroken.
Our classical configurations, however, should not be extrapolated back to
times of the order of the Planck scale. The ``stringy'' dynamics of the
universe at or before 
the Planck epoch is a subject of current interest \cite{Jmu, PREBB}.
Work on quantum effects within the overall scheme of supersymmetric 
compactification is in progress.

\vspace{0.5cm}
One objection to our approach could be that supersymmetry consistency
conditions are not, after all, equations of motion.
Can we, without invoking the latter, make definite statements about
dynamics?  
Our answer is yes.
The uncertainties inherent in our approach are no greater than the
uncertainty regarding the form of the dilaton potential in other
approaches \cite{Brustein, OliveNPB, OlivePLB, Jmu}.
We can go a step further, and conjecture that the strong ``equation of motion''
look of the consistency conditions is not accidental.
There is a deep dynamical content in these equations which, we
believe, will be discovered at some time in the future.

%% file: ack.tex
This work is part of a project (No. SP/S2/K--06/91) funded by the
Department of Science and Technology, Government of India.  
H.K.J. thanks the University Grants Commission, New Delhi, for a fellowship.  
The authors are grateful to Sayan Kar for pointing out an error in an earlier version of this paper.

%% file: figure.tex

\begin{enumerate}

\item Evolution of the scale factor ref. \cite{MMS}.
\vspace{0.5cm}
\item Evolution of the dilaton field ref. \cite{MMS}.
\vspace{0.5cm}
\item The scale factor $R(t)$---new solutions.
\vspace{0.5cm}
\item The dilaton field $\phi(t)$---new solutions. 
\vspace{0.5cm}
\item Solutions to the dual equations---$R(u)$ vs $u$.
\vspace{0.5cm}
\item Solutions to the dual equations---$\chi(u)$ vs $u$.
\vspace{0.5cm}
\item Solutions to the dual equations---$R(t)$ vs $t$.
\vspace{0.5cm}
\item Solutions to the dual equations---$\chi(t)$ vs $t^2$.
\vspace{0.5cm}
\item  Solutions to the dual equations---$\phi$ vs $u$.
\vspace{0.5cm}
\item Solutions to the dual equations---$\phi(t)$ vs $t$.

\end{enumerate}